\documentclass[prb,twocolumn,showpacs,preprintnumbers,amsmath,amssymb,superscriptaddress]{revtex4}
\usepackage{graphicx}
\usepackage{dcolumn}
\usepackage{bm}

\begin{document}
\title{Angular-dependent Magnetoresistance Oscillations in Na$_{0.48}$CoO$_{2}$ Single Crystal }

\author{F. Hu}
\affiliation{Institute of Theoretical Physics and
Interdisciplinary Center of Theoretic Studies, Chinese Academy of
Sciences, P.O. Box 2735, Beijing 100080, China}

\author{G. T. Liu}
\affiliation{Beijing National Laboratory for Condensed Matter
physics,Institute of Physics, Chinese Academy of Science, P. O.
Box 603, Beijing 100080, China}

\author{J. L. Luo}
\email{JLLuo@aphy.iphy.ac.cn} \affiliation{Beijing National
Laboratory for Condensed Matter physics,Institute of Physics,
Chinese Academy of Science, P. O. Box 603, Beijing 100080, China}

\author{D. Wu}
\affiliation{Beijing National Laboratory for Condensed Matter
physics,Institute of Physics, Chinese Academy of Science, P. O.
Box 603, Beijing 100080, China}

\author{N. L. Wang}
\affiliation{Beijing National Laboratory for Condensed Matter
physics,Institute of Physics, Chinese Academy of Science, P. O.
Box 603, Beijing 100080, China}

\author{T. Xiang}
\affiliation{Institute of Theoretical Physics and
Interdisciplinary Center of Theoretic Studies, Chinese Academy of
Sciences, P.O. Box 2735, Beijing 100080, China}
\affiliation{Center for Advanced Study, Tsinghua University,
Beijing 100084, China}

\begin{abstract}

We report measurements of the c-axis angular-dependent
magnetoresistance (AMR) for a Na$_{0.48}$CoO$_{2}$ single crystal,
with a magnetic field of 10 T rotating within Co-O planes. Below
the metal-insulator transition temperature induced by the charge
ordering, the oscillation of the AMR is dominated by a two-fold
rotational symmetry. The amplitudes of the oscillation
corresponding to the four- and six-fold rotational symmetries are
distinctive in low temperatures, but they merge into the
background simultaneously at about 25 K. The six-fold oscillation
originates naturally from the lattice symmetry. The observation of
the four-fold rotational symmetry is consistent with the picture
proposed by Choy $\emph et\hspace{0.2cm}al.$ that the Co lattice
in the charge ordered state will split into two orthorhombic
sublattice with one occupied by Co$^{3+}$ ions and the other by
Co$^{4+}$ ions. We have also measured the c-axis AMR for
Na$_{0.35}$CoO$_{2}$ and Na$_{0.85}$CoO$_{2}$ single crystals, and
found no evidence for the existence of two- and four-fold
symmetries.

\end{abstract}

\date{\today}
\maketitle

Recently great attention has been paid to the investigation of
physical properties of the newly discovered superconductor
Na$_{0.35}$CoO$_{2} \cdot {1.3}$H$_{2}$O with $\it T_{c}\sim
4.5$K.\cite{Takata} Until now it is the only known layered
transition metal oxide which exhibits superconductivity other than
high-$T_{c}$ cuprates and Sr$_{2}$RuO$_{4}$. The non-hydrated host
oxide Na$_{x}$CoO$_{2}$ is known for its large thermoelectric
power.\cite{I.Terasaki,Yayu Wang} It is a two-dimensional cobalt
oxide. Co atoms form a triangular lattice rather than a square one
as in high-$T_{c}$ cuprates and the Co-O planes are composed of
the edge sharing O-octahedra with a Cobalt atom lying in the
center of every O-octahedron. Na atoms or H$_{2}$O molecules in
the superconductor lie between two neighboring Co-O
planes.\cite{Q.Huang 1,D.P.Chen}

The phase diagram of Na$_{x}$CoO$_{2}$ shows that this cobalt
oxide is a paramagnetic metal at $x \sim 0.35$, a Curie-Weisse
metal around $x\sim 0.66$, and a magnetically ordered metal at
$x\sim 0.85$. However, at about $x=0.5$, Na$_{x}$CoO$_{2}$ becomes
a charge-ordered insulator in low
temperatures.\cite{M.L.Foo,J.L.Luo} It was reported that
Na$_{0.5}$CoO$_{2}$ exhibits three successive phase transitions at
88 K, 53 K and 20 K, respectively.\cite{M.L.Foo,Q.Huang
1,H.X.Yang} A charge ordered state is formed in low temperatures.
It causes a metal-insulator transition at 53 K. Furthermore, a
static magnetic order was observed by the $\mu$SR measurements
below this metal-insulator transition
temperature.\cite{Y.J.Uemura} A weak magnetic transition was
observed in the resistivity as well as low field (a few gausses)
magnetic susceptibility measurements at 25 K.\cite{H.X.Yang} The
$\mu$SR experiments also showed that there are two characteristic
frequencies below 20 K, indicating the formation of two distinct
types of local moments.

Recently, the c-axis magnetoresistance of Na$_{0.5}$CoO$_{2}$ was
measured.\cite{L.Balicas,C.H.Wang} It was found that a magnetic
field applied along the c-axis direction has almost no effect on
the metal-insulator transition temperature up to 43 Tesla. On the
contrary, if a magnetic field is applied along the Co-O planes,
the charge ordering is suppressed and the metal-insulator
transition disappears above 40 Tesla. The c-axis AMR measurement
\cite{L.Balicas} indicates that the electronic structure changes
from a two-fold to a six-fold symmetry when the magnetic field
rotating within the Co-O planes is increased from 25 T to 45 T at
0.6 K.

In this paper, we report the measurement of the c-axis AMR for
Na$_{0.48}$CoO$_{2}$ single crystal, with a rotating magnetic
field of 10 Tesla applied within Co-O planes. As will be shown
later, the AMR measurement is a useful tool for analyzing
electronic states of materials with layered structures. It can
reveal, for example, the symmetry of the Fermi surface as well as
the distribution of electronically active
ions.\cite{L.Balicas,N.E.Hussey}

The Na$_{0.48}$CoO$_{2}$ sample is prepared by sodium
deintercalation from a Na$_{0.85}$CoO$_{2}$ single phased single
crystal, synthesized by the floating zoom
method.\cite{J.L.Luo,D.Wu} The single crystal of
Na$_{0.85}$CoO$_{2}$ was marinated in a Br$_{2}$ and CH$_{3}$CN
mixture solution for more than two weeks for sodium
deintercalation. Sodium concentration determined by ICP is
$0.48\pm0.02$. X-ray measurements show that Na$_{0.48}$CoO$_{2}$
is in a single phase and the c-axis lattice constant $c$ is about
11.04 \AA. The magnetoresistance is measured using a standard
four-probe low frequency ac method  in a PPMS system of Quantum
Design company.

\begin{figure}[ht]
\includegraphics[width=6.5cm]{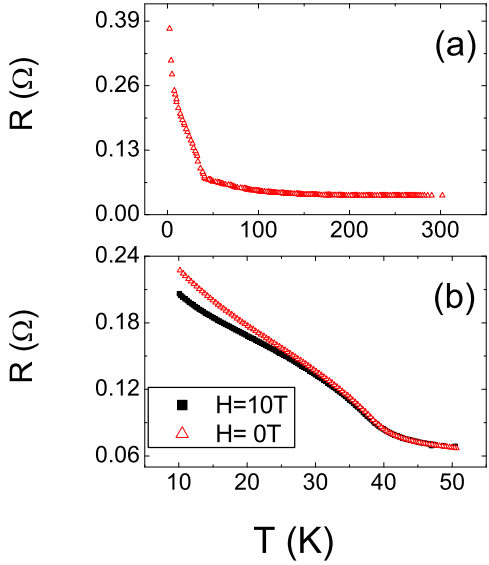}
\vspace{-0.2cm} \caption{(a) Temperature dependence of the c-axis
resistivity $R$ of the Na$_{0.48}$CoO$_{2}$ single crystal. (b)
$R$ versus $T$ below 50 K at 0 T (open triangles) and 10T applied
within the Co-O planes (solid squares), respectively.}
\end{figure}

Figure 1(a) shows the c-axis resistance $R$ of the
Na$_{0.48}$CoO$_{2}$ single crystal as a function of temperature
$T$. The overall $R-T$ curve behaves similarly as for
Na$_{0.5}$CoO$_{2}$.\cite{M.L.Foo,Q.Huang 1} With decreasing
temperature, $R$ increases gradually from 300 K to 40 K. It then
shows a dramatic increase below 40 K, as a result of
metal-insulator phase transition. Here the metal-insulator
transition temperature is lower than the corresponding transition
temperature for Na$_{0.5}$CoO$_{2}$. This difference is due to the
difference in the Na concentration.

Figure 1(b) compares the temperature dependence of $R$ at zero
field with that in an external magnetic field of 10 T applied
along the Co-O planes. It shows that the metal-insulator
transition temperature is not changed by the applied field. A
positive magnetoresistance, which increases with decreasing
temperature, is observed below 40 K. $R$ exhibits a weak hump
structure between 10 K and 30 K. In Na$_{0.5}$CoO$_{2}$, there
exists a similar hump in the $R-T$ curve between 25 K and 40 K.

\begin{figure}[ht]
\includegraphics[width=8.5cm]{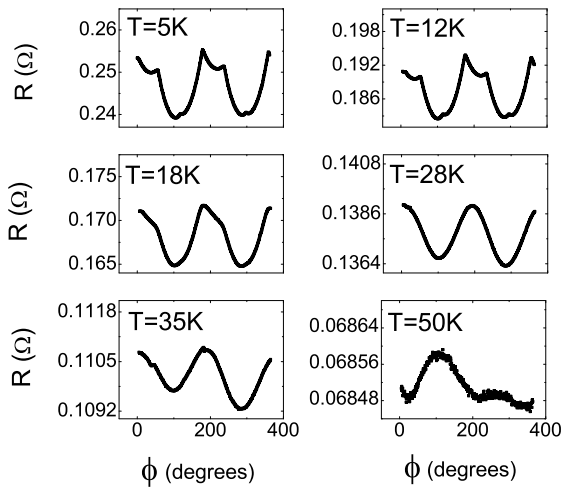}
\vspace{-0.2cm} \caption{Angular dependence of the c-axis
resistance of Na$_{0.48}$CoO$_{2}$ single crystal at different
temperatures in an applied magnetic field of 10T rotating within
Co-O planes.}
\end{figure}

Figure 2 shows the c-axis AMR for Na$_{0.48}$CoO$_{2}$ at
different temperatures. In low temperatures, the AMR shows a clear
periodical oscillation decorated with fine structures with the
change of the rotating angle $\phi$. However, in high
temperatures, some of these fine structures disappear.

\begin{figure}[ht]
\includegraphics[width=7cm]{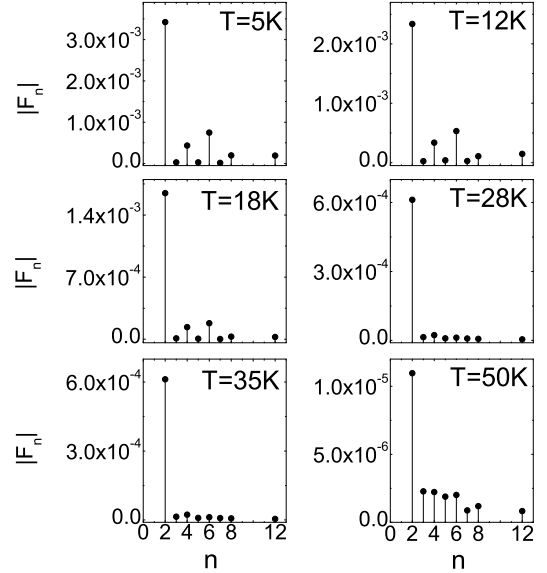}
\vspace{-0.2cm} \caption{The Fourier power spectra of the c-axis
AMR for Na$_{0.48}$CoO$_{2}$ at different temperatures.}
\end{figure}

In order to find the hidden periodicities associated with the fine
structures of the AMR, we perform a Fourier analysis on the
experimental data. The Fourier transformation of $R(\phi )$ is
defined by
\begin{equation}
F_n=\frac{1}{N} \sum_{j=0}^{N-1} R(\phi_{j}) e^{-i\phi_{j}  n}
\end{equation}
where $\phi_{j}= 2\pi j/ N$ and $n=0,1,\cdots N-1$. $N$
$(\sim450)$ is the total number of data points measured at each
circle of rotation of the magnetic field. $F_n$ is the $n$'th
coefficient of the Fourier expansion. Its absolute value $|F_n|$
measures the contribution of the conduction channel with $n$-fold
rotational symmetry.

Figure 3 shows the Fourier power spectra $|F_n|$ at different
temperatures from 5 K to 50 K. Here we concentrate on the
components that are physically most relevant, i.e. $n=2\sim8$ and
12. We find that $F_2$, $F_4$ and $F_6$ are the three most
important components. Among them, the two-fold symmetric component
($n=2$) is the most distinctive one. This indicates that the
two-fold symmetric channel contributes most to the c-axis
magnetoresistance below 50 K. This is apparently due to the
formation of charge ordering in this temperature regime. In
Na$_{x}$CoO$_{2}$ there are both Co$^{3+}$ and Co$^{4+}$ ions. In
the charge ordered state, it was believed that Co$^{3+}$ and
Co$^{4+}$ ions will order independently to form alternating charge
stripes.\cite{M.L.Foo,Q.Huang 1,H.W.Zandbergen} These charge
stripes reduce the symmetry of the material and lead to the $F_2$
term in the c-axis AMR. Similar two-fold symmetry has been
observed in the in-plane as well as out-plane AMR in lightly doped
high-$T_{c}$ cuprates La$_{2-x}$r$_x$CuO$_4$\cite{Ando}. In that
material, the two-fold symmetry is believed to be due to the
formation of charge stripes. The six-fold symmetry in AMR reflects
naturally the triangular lattice symmetry of Na$_{x}$CoO$_{2}$. It
is also a manifestation of the six-fold rotational symmetry of the
Fermi surface of Na$_{x}$CoO$_{2}$ as revealed by the
Angle-Resolved Photoemission measurements\cite{M.Z.Hasan,Yang}.

\begin{figure}[ht]
\includegraphics[width=6cm]{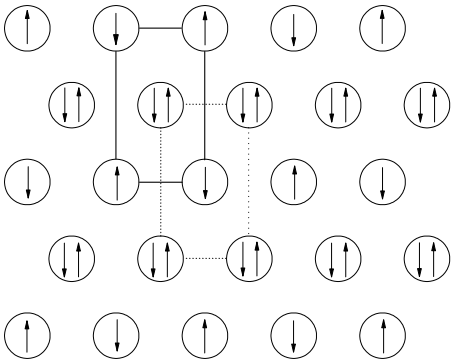}
\vspace{-0.2cm} \caption{ A possible low temperature structure of
the charge-ordering state of Na$_{0.5}$CoO$_{2}$. The circles
represent cobalt ions. The triangular lattice is broken into two
orthorhombic sublattices with one occupied by Co$^{3+}$ ions and
the other by Co$^{4+}$ ions with antiferromagnetic ordering. }
\end{figure}

The observation of the distinctive four-fold symmetric component
in low temperatures indicates that there is a square-lattice-like
symmetry in Na$_{0.48}$CoO$_{2}$. In Na$_{0.5}$CoO$_{2}$, neutron
and electron diffraction experiments revealed that Na atoms can
form an orthorhombic superstructure with a=2a$_{h}$,
b=$\sqrt{3}$a$_{h}$.\cite{Q.Huang 1,H.W.Zandbergen} The ordering
of Na atoms can lead to an ordering state of charge carriers in
the Co-O layers. \cite{Q.Huang 1,P.Zhang} Recently, Choy $\emph
et\hspace{0.2cm}al.$ proposed an insulating ground state of
Na$_{0.5}$CoO$_{2}$ based on the calculation of the Hubbard model.
As shown in figure 4, the triangular Co lattice is broken into two
orthorhombic sublattices with one occupied by Co$^{3+}$ ions and
the other by Co$^{4+}$ ions.\cite{Choy} The Co$^{4+}$ ions, each
of which contains a local moment with S=1/2, form an
antiferromagnetic sublattice with dimension $a \times\sqrt{3}a$.
The $\mu$SR and magnetic susceptibility experiments also suggest
that there exists a spin long range order below 53 K, and this
spin long range order may be antiferromagnetic
correlated.\cite{Y.J.Uemura,M.L.Foo} The observation of the
four-fold rotational symmetry in the AMR lends support to this
picture.

\begin{figure}[ht]
\includegraphics[width=6.5cm]{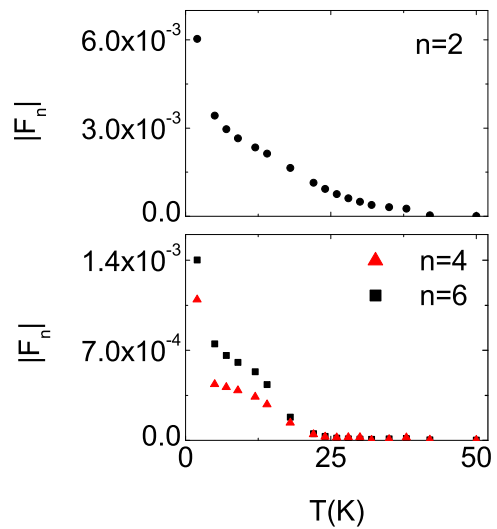}
\vspace{-0.2cm} \caption{The temperature dependence of $|F_2|$,
$|F_4|$ and $|F_6|$ for Na$_{0.48}$CoO$_{2}$.}
\end{figure}

Figure 5 shows the temperature dependence of $|F_2|$, $|F_4|$ and
$|F_6|$. $|F_2|$ decreases with increasing temperature and
vanishes at roughly the charge-ordering temperature ($\sim 40 K$).
This suggests that the two-fold symmetry in the AMR is indeed due
to the charge ordering in Na$_{0.48}$CoO$_{2}$. With increasing
temperature, both $|F_4|$ and $|F_6|$ drop quickly in low
temperatures and then merge into the noise background at about 25
K.

\begin{figure}[ht]
\includegraphics[width=9cm]{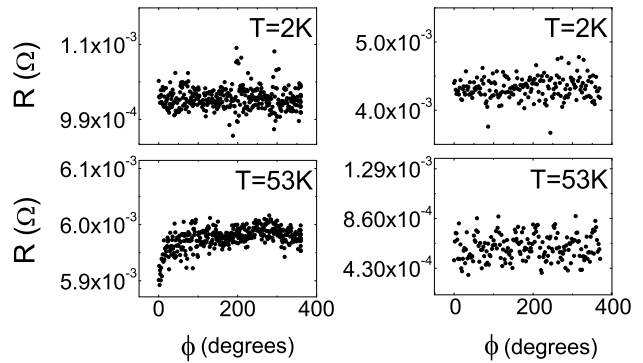}
\vspace{-0.2cm} \caption{ The c-axis AMR for Na$_{0.35}$CoO$_{2}$
(left panel) and  Na$_{0.85}$CoO$_{2}$ (right panel) single
crystals.}
\end{figure}

It is interesting to compare the AMR results for
Na$_{0.48}$CoO$_{2}$ with the compounds that do not show any
charge ordering phase in low temperatures. Figure 6 shows the
c-axis AMR results for Na$_{0.35}$CoO$_{2}$ and
Na$_{0.85}$CoO$_{2}$ single crystals. In contrast to
Na$_{0.48}$CoO$_{2}$, the c-axis AMR for Na$_{0.35}$CoO$_{2}$ and
Na$_{0.85}$CoO$_{2}$ does not show any periodical oscillations
above the noise background. It is not surprising that the two- and
four-fold symmetric components are absent in the AMR, since there
is no charge ordering of Co and Na ions in both
Na$_{0.35}$CoO$_{2}$ and Na$_{0.85}$CoO$_{2}$. However, the
absence of six-fold symmetry is probably due to the fact that the
c-axis AMR in Na$_{0.35}$CoO$_{2}$ and Na$_{0.85}$CoO$_{2}$ are
significantly smaller than that of Na$_{0.48}$CoO$_2$.

In summary, we have measured the c-axis AMR of a
Na$_{0.48}$CoO$_{2}$ single crystal. We find that the oscillation
of the AMR is dominated by a two-fold rotational symmetry below
the metal-insulator transition temperature. This is consistent
with the observation that the metal-insulator transition is
induced by the charge ordering of Co ions. The amplitudes of the
oscillation in the AMR corresponding to the four- and six-fold
rotational symmetries are also distinctive, but disappear at about
25 K. The six-fold oscillation reflects naturally the triangular
lattice symmetry. The four-fold oscillation is speculated to
result from the orthorhombic redistribution of Co ions in low
temperatures.

This work was supported by the National Natural Science Foundation
of China.

\end{document}